\documentclass[a4paper,11pt]{article}
\pdfoutput=1 

\usepackage{jheppub} 

\title{A Low Dead Time, Resource efficient Encoding Method for FPGA based High-Resolution TDL TDCs}

\author[a,b]{W.Dong,}
\author[a,b,1]{C. Feng,\note{Corresponding author.}}
\author[a,b]{J. Wang,}
\author[a,b]{Z. Shen,}
\author[a,b]{S. Liu,}
\author[a,b]{and Q. An}

\affiliation[a]{State Key Laboratory of Particle Detection and Electronics, University of Science and Technology of China,\\Hefei 230026, China}
\affiliation[b]{Department of Modern Physics, University of Science and Technology of China,\\Hefei 230026, China}

\emailAdd{fengcq@ustc.edu.cn}

\abstract{This paper presents a novel encoding method for fine time data of a tapped delay line (TDL) time-to-digital Converter (TDC). It is based on divide-and-conquer strategy, and has the advantage of significantly reducing logic resource utilization while retaining low dead-time performance. Furthermore, the problem of high bubble depth in advanced devices can be resolved with this method. Four examples are demonstrated, which were implemented in a Xilinx Artix-7  Field Programmable Gate Array (FPGA) device, and encoding method presented in this paper was employed to encode fine time data for normal TDL TDC, a half-length delay line TDC, and double-edge and four-edge wave union TDCs. Compared with TDCs from the latest published papers that adopt traditional encoders, the logic utilization of TDCs in this paper were reduced by a factor of 45\% to 70\% in different situations, while the encoding dead time can be restricted in one clock cycle.  Acceptable resolutions of the  demonstrated TDCs were also obtained, proving the functionality of the encoding method.}


\begin{document}
\maketitle
\flushbottom

\section{Introduction}
\label{sec:intro}

Time-to-digital converter (TDC) plays a key role in many areas, such as particle physics experiments, medical imaging equipment, and autopilot systems. For example, a time-of-flight (TOF) detector is often the core part of large detector systems for particle and astroparticle physics projects~\cite{carnesecchi2019performance}. Besides, TOF detectors determine the coincidence time of positron emission tomography imaging, the range of a LIDAR receiver, and the distance of direct TOF cameras. All of these applications require high-resolution TDCs ~\cite{brunner2013studies, keel2019640, song2017high}. 

Time-to-digital converters are mainly implemented in application-specific integrated circuits (ASICs) or Field Programmable Gate Array (FPGA) in recent years. 
Benefiting from their advantages in flexibility and development cycle, FPGA-based TDCs have been widely applied in many areas where the requirements of TDCs are non-standard and not massive. In recent years, FPGA-based TDCs have experienced remarkable advances, and their accuracy has reached the level of several pico-seconds, which is comparable to the performance of state-of-the-art ASICs, mainly due to the invention of dedicated carry chain~\cite{song2006high, wu2003firmware} and wave union architectures~\cite{wu200810, wang201110}.

However, increasing the number of channels in a single chip is still a challenge. The density of TDC channels has become increasingly important in various applications as mentioned above. For example, in the ALICE (A Large Ion Collider Experiment) experiment at the Large Hadron Collider, there were 152,928 readout channels in its TOF detector, and more than 20 thousand high-resolution TDC ASIC chips were used, with eight channels for each chip~\cite{carnesecchi2019performance}. In another example of a Compressed Baryonic Matter spectrometer, the TOF subsystem contained thousands of readout channels (1,600 of mTOF and 6,912 of eTOF)~\cite{deppner2020fair}.  For applications that contain a large amount of time measurement channels, increasing the number of channels in a single TDC chip is a direct way to improve the system integration level. However, the logic utilization becomes a pivotal factor that restricts the channel density for FPGA-based TDCs. Logic resources are limited in FPGAs and usually other functions, than the core TDC, are required in the same chip.

Among all the architectures, a normal tapped delay line (TDL) is a mainstream type for FPGA-based TDCs. It combines a coarse counter that is driven by the system clock and a TDL for fine-time interpolation. The latter is usually implemented by a carry chain in FPGA. In this way, a large dynamic range together with a high time resolution is achieved. However, to obtain the fine time which is less than one clock period, the TDL's raw data must be encoded into binary code. Since the raw data is usually thermometer code, some encoders used in flash ADC have been used in this kind of TDC~\cite{zheng2017low}. 

For normal FPGA-based TDL TDC, the time resolution is limited by the cell delay. Some methods have been employed to attain higher precision, among which multi-chain architecture is an effective way to improve the time resolution beyond its cell delay. However, this method occupies more logic resources not only to implement more carry chains but also to encode the fine time data for all the chains. Therefore, it is not suitable for applications that need many TDC channels.

Wave Union, proposed by J. Wu and Z. Shi ~\cite{wu200810}, is another effective architecture to achieve higher precision. It employs two types of "Wave union launchers" (named A and B) to make multiple measurements with a single delay chain structure, which effectively sub-divides the ultra-wide bins in each raw measurement. For the Wave Union A structure, which is more widely used because of its low dead time feature, the launcher generates one or several narrow pulses and all the rising and falling edges are sampled in one clock cycle. In this case, raw data of the TDL becomes non-thermometer code. 

A type of newly presented architecture~\cite{kong2020resource}, referred to as half-length delay line TDC in this paper, has this type of non-thermometer raw data as well. For each effective input signal, a square pulse, with the width of roughly half a clock cycle, is generated, and any one of its two edges is captured to produce a fine timestamp. In this way, the length of the TDL can be shorten to slightly longer than half of the clock cycle. 

However, for the above two TDC architectures, the nonthermometer-to-binary (NTH2B) encoding logic is a particular challenge because of the variable and complex raw data pattern. To deal with this problem, several methods, such as a stepped-up tree encoder, have been presented ~\cite{hu2013stepped}. However, they are much less resource efficient and less scalable than normal thermometer-to-binary (TH2B) encoders, which has become restrictions for further increasing the channel density.

In this paper, a novel resource-saving encoding method is proposed, which can be used in both thermometer-to-binary and non-thermometer-to-binary encoders. It is based on the strategy of divide-and-conquer, which is fully parallel in time and capable of correcting bubble errors. Based on this method, high channel density TDCs can be realized in a low cost FPGA chip.

\section{Methodology and design concept}

\subsection{Basic concepts}

\begin{figure}[htb]
\centering 
\includegraphics[scale=3]{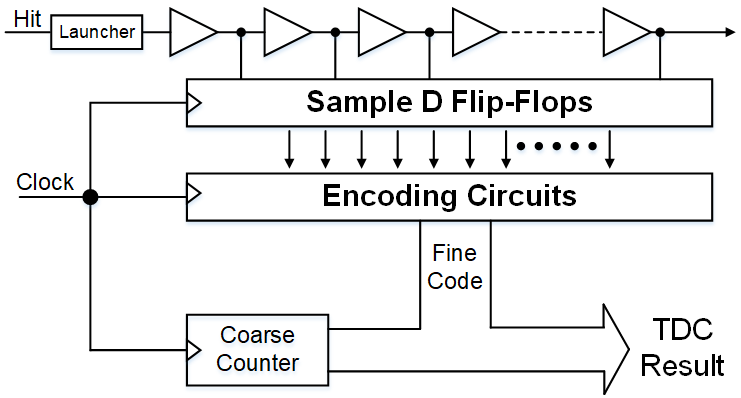}
\caption{\label{fig:1} Basic components of TDL TDCs.}
\end{figure}

The architecture of a typical TDL TDC is shown in Figure \ref{fig:1}. Besides a coarse counter that runs at the system clock, it is mainly composed of a TDL, an array of D flip-flops, and the encoding circuits. A specific waveform is generated and transmitted on the delay-line when the launcher is triggered by an input signal (Hit). The outputs of all the delay cells of the TDL are sampled in the form of raw data by D flip-flops on every rising edge of the system clock. Finally, the raw data of the TDL is encoded. If the raw data indicates that a transient waveform appears on the delay line, then the output of the encoding circuits and the coarse counter is buffered subsequently by auxiliary circuits. 

In this paper, three types of TDCs are discussed, which have the same kind of delay line, sample D flip-flop, and coarse counter, but the waveforms running on the TDL are different. The three types of TDCs are 1) step signal for a normal TDL-TDC, 2) spike pulse for a wave union TDC, and 3) a square wave for a half-length delay line TDC. The wave union TDC discussed in this paper is type A, which is measured in one clock cycle~\cite{wu200810}. Therefore, the launchers and encoders in the three types of TDCs are also different. The encoder for a half-length delay line TDC needs to deal with raw data corresponding to either the rising or falling edge, while the encoder for a wave union TDC has to deal with more than one edge.

\begin{figure}[htb]
\centering 
\includegraphics[scale=0.6]{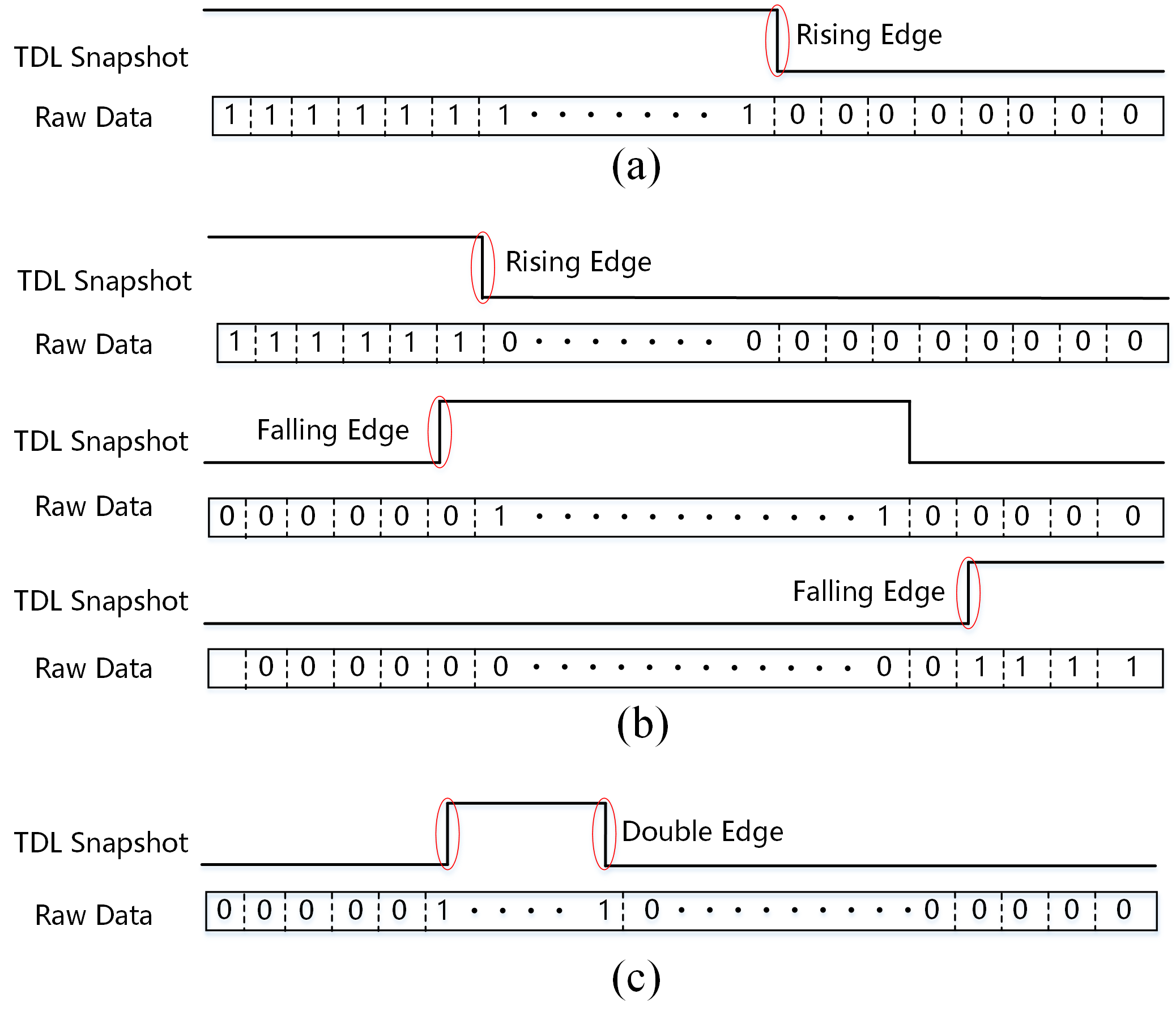}
\caption{\label{fig:2} Signal propagations and the delay chains’ data patterns of three kinds of TDCs: (a) a normal TDL TDC (the rising edge is captured ); (b) a half-length delay line TDC (either rising or falling edge is used depending on the status of the propagation in the TDL); and (c) a wave union TDC (both the rising and falling edges are captured at the same time).}
\end{figure}

Figure \ref{fig:2} shows the three kinds of snapshots when hit signals propagate on delay line and their corresponding raw data. The pattern of the raw data of a normal TDL TDC is in the form of “111…1100...000,” and the pattern of a wave union TDC is in the form of “00…0011…1100…00.” 

As for half-length delay line TDCs, a short delay line is used, which covers just slightly more than one-half of the clock cycle. There are three kinds of patterns, “111…1100…000,” “000…0011…1100…000,” and “000…0011…111,” which correspond to different statuses of signal propagation. Although there are various types of input data, the output data of the encoder is binary code, which contains the positions of 1-to-0 or 0-to-1 transitions. These transitions indicate the edges of the input signal which are captured by the delay line and DFFs. 

The basic idea of all the encoding methods is to find the edge. A simple method is to employ the exclusive-OR (XOR) logical operation for every two adjacent binary codes of the raw data. Then, the edge position is encoded directly by a priority encoder. However, this method is based on the ideal raw data which is in form of “111…1100…00”. In advanced devices, researchers have found that the raw data sometimes have the pattern like “111…1100110100…000”, in which the zeros between ones are called “bubbles”~\cite{wu2010several}. The length of the blurring area, such as “001101”, is defined as bubble depth. And the simple XOR method can’t work in this situation.

The exact source of bubbles is difficult to find out for FPGA users. But considering the basic principles of digital circuits, the non-ideality of clock distribution and placement of logic cell may be the main factor. In addition, the signals propagating in the TDL are not synchronized to the sampling clock. So metastability may be another root of bubble problem.

In order to solve this problem, the mainstream methods are to count the “1”s in raw data, among which the Wallace tree encoder is a mature technology that has been applied widely in normal TDC~\cite{zheng2017low}. Some other encoders that were originally invented for Flash ADC, such as Ones-Counter, have also been introduced to normal TDL TDC~\cite{sall2007thermometer,wang20173}, but most of them are not capable of handling non-thermometer-to-binary codes. Another encoder was proposed to eliminate the bubble problems by using the first 1-0 or 0-1 transition ~\cite{jiao2021resource}. However, it is based on the premise that the maximum bubble depth is four, and the encoder structure is difficult to be extended for the situations of deeper bubbles. 

The encoding method proposed in this paper is based on divide-and-conquer strategy. It solves the bubble problems by dividing the raw data into pieces and counting "1"s in each piece.
The number of "1"s in each piece is called "local sum" and the MSB of the local sum is the flag of the corresponding piece. The flags of these pieces are used to find the one or more transitions in the raw data roughly. For each transition, two adjacent pieces where the transition locates in are selected . Finally, combined with the local sums of the two selected pieces, the exact position of the transition is attained.

\subsection{Architecture of encoding circuits}

The encoding method proposed in this paper has several features: resource saving, eliminating bubbles and an extensible and standard structure for different types of TDL TDCs. The encoding method has three steps as follows: The first step is to divide the TDC raw data into small pieces and then the local sums and flags are attained. By making use of the flags from the first step, the second step is to find the edges piece by piece. In the final step, the edge position is calculated from the sequence number of the selected piece of the second step and the local sums of the first step.  

\begin{figure}[htb]
\centering 
\includegraphics[scale=0.50]{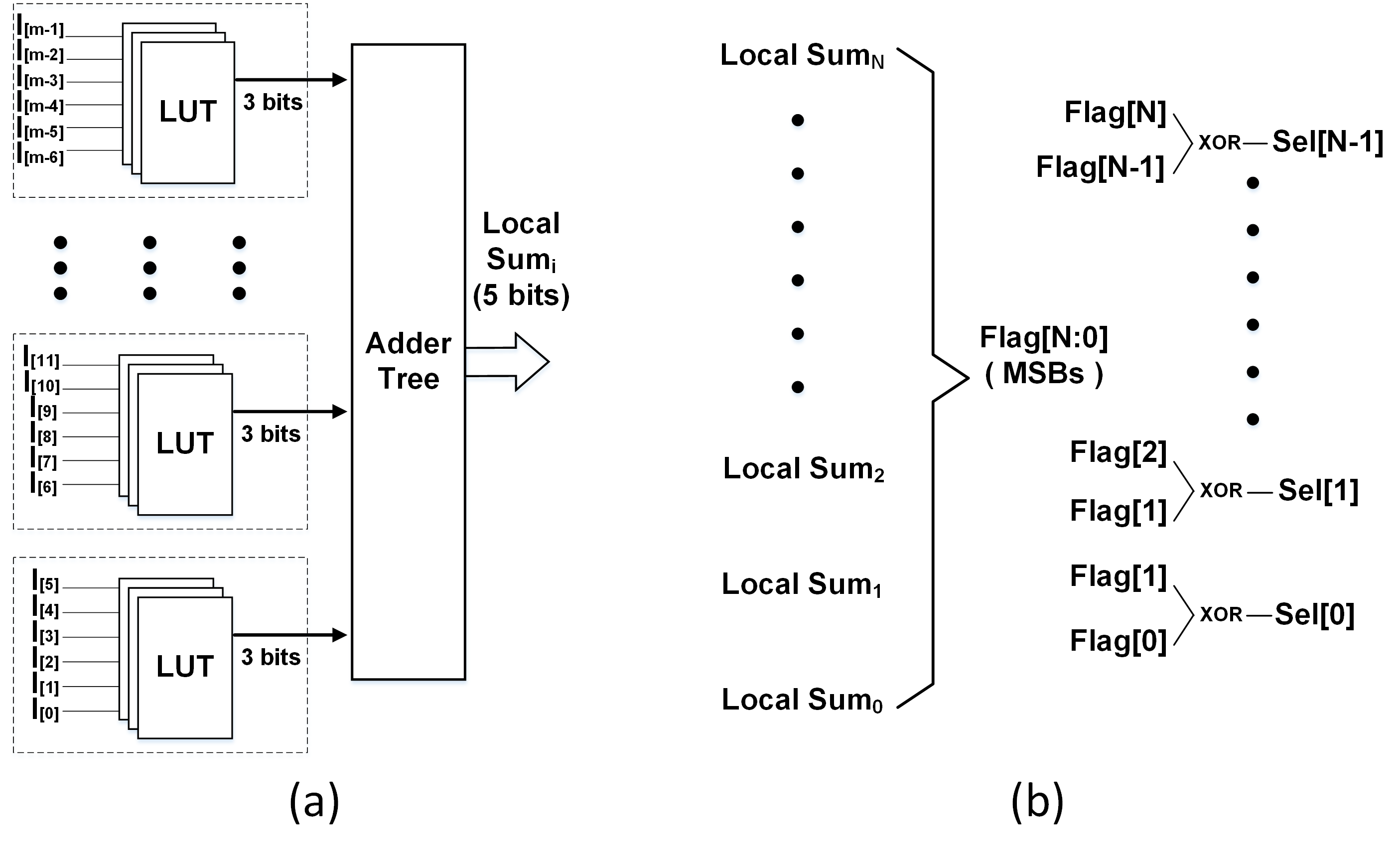}
\caption{\label{fig:4} The process of first step in the encoding method. It consists of pre-encoded cells and gets labels that roughly show the edge’s position. a) The details of pre-encoded cell; b) the process of  getting the Sel array.}
\end{figure}

In the first step, the raw bins are divided into m-bit (m equals 24 in Figure \ref{fig:4}) wide pieces. The pre-encoded cells get the local sums of these pieces. After being buffered by D Flip-Flops, the pre-encoded results are sent to back-end circuits to find the 1-0 transition. As shown in Figure \ref{fig:4}, the flag array consists of the most significant bits (MSB) of each local sum and the piece select (Sel) array is composed of the XOR results of the adjacent bits of the flag array. 

\begin{figure}[htb]
\centering 
\includegraphics[scale=0.53]{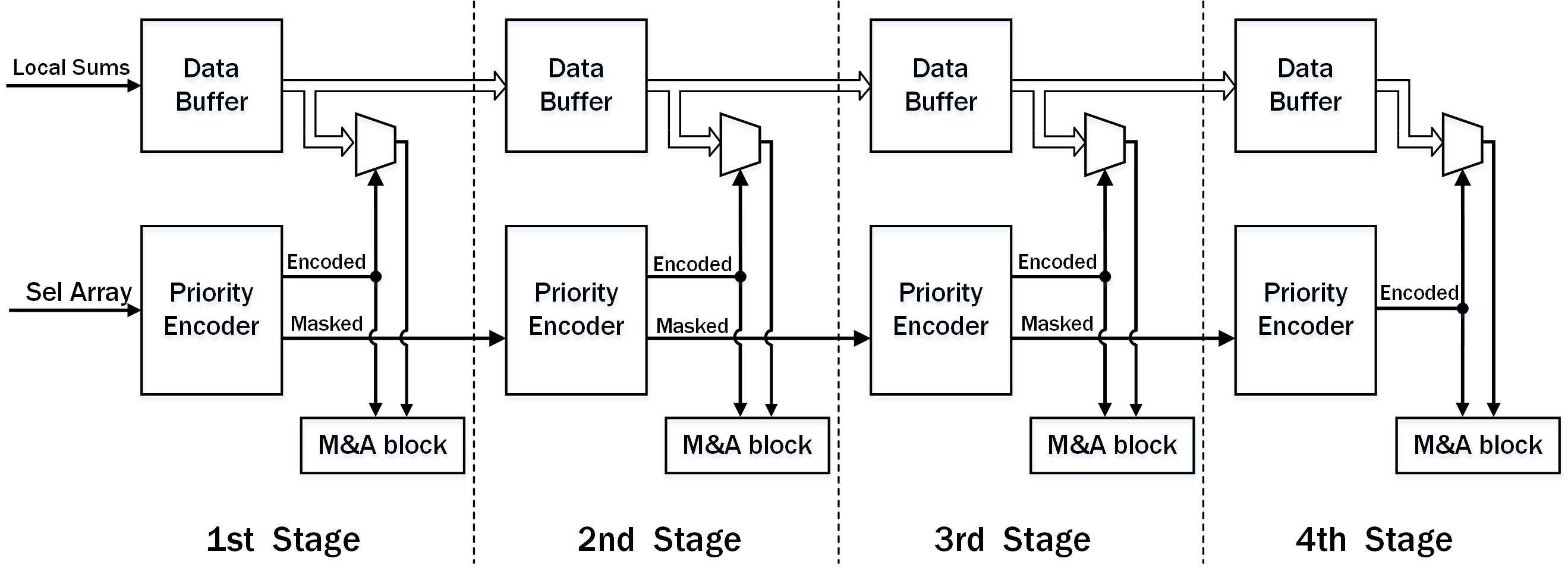}
\caption{\label{fig:5} The back-end encoding circuits consist of priority encoders, multiplexers, and arithmetic circuits(the M\&A block). The diagram shows encoders for four-edge wave union TDCs.}
\end{figure}

In the second step, the encoder finds which two adjacent pieces the transition occurs in. The output of the first step is buffered. The back-end circuit, which accomplishes the second and final steps, consists of  priority encoders, multiplexers, and arithmetic circuits. The back-end circuit finds pieces that have a 1-0 or 0-1 transition by the Sel array in the priority encoder. Finally, the product of the encoding result and the piece width plus the local are summed in the arithmetic circuits. The arithmetic circuits consist of multipliers and adders. The binary code of one transition's location is attained. The back-end circuit for four-edge wave union TDCs is designed to be a pipeline structure as shown in Figure \ref{fig:5}. 

In further steps, the difference between the encoding method and the Wallace tree encoder is that the priority encoders and multiplexers in this method replace the Wallace tree encoder’s adders. The input of the MUXs is not the most significant bit (MSB) of the sum of these pieces but the XOR of their adjacent MSB (as the Sel array shows in Figure \ref{fig:4}). As a result, compared with the Wallace tree encoder, the encoding method presented in this paper can deal with non-thermometer-code. As shown in Figure  \ref{fig:5}, the encoding method can deal with multiple transitions by cascading priority encoders.

\subsection{Three types of encoders}

\subsubsection{Encoder for a normal TDL TDC}

\begin{figure}[htb]
\centering 
\includegraphics[scale=0.5]{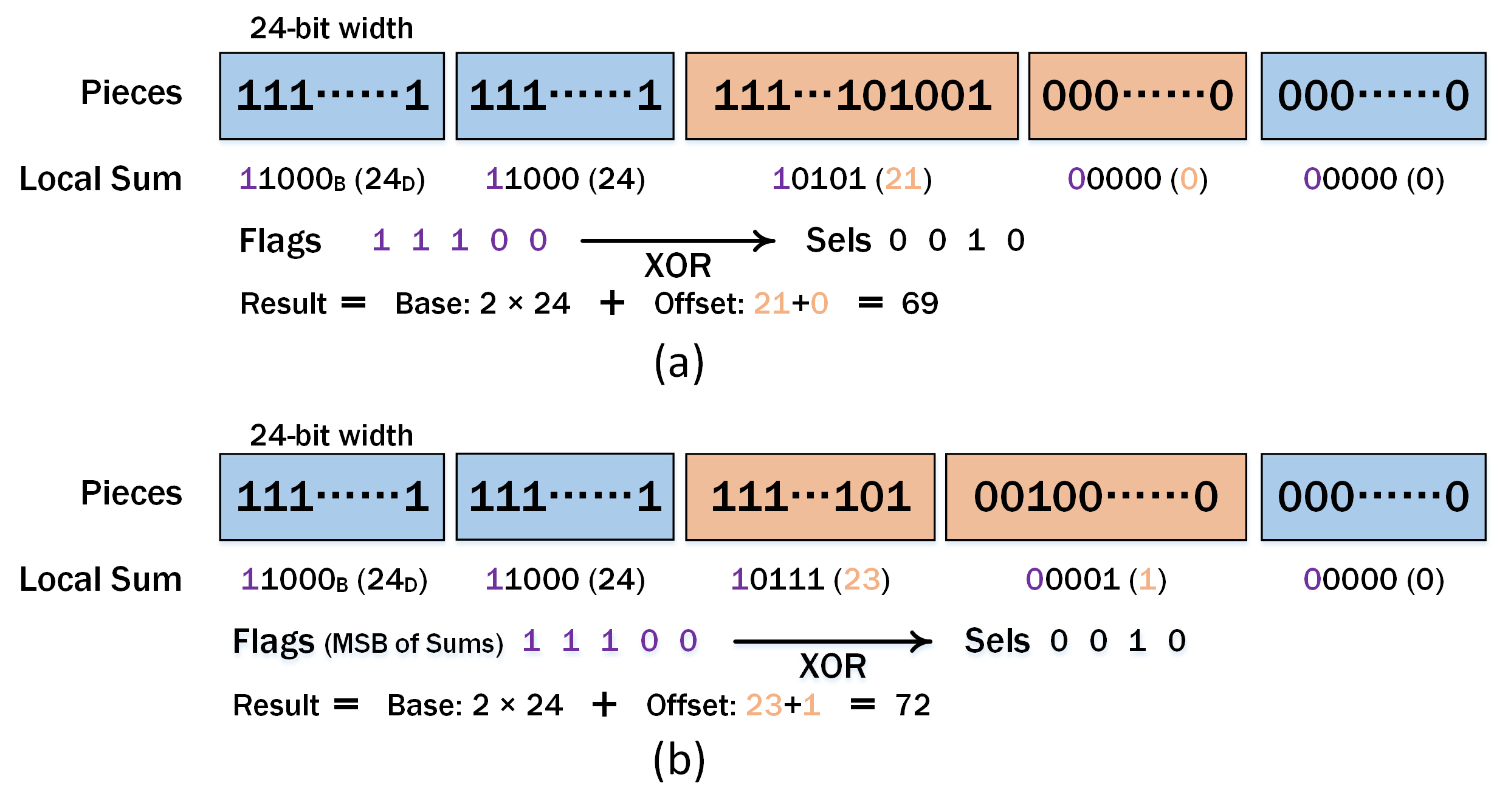}
\caption{\label{fig:3} Examples of calculating transition position in Normal TDL TDCs.}
\end{figure}

Normal TDL TDCs only need to encode one 1-0 transition’s position, which is illustrated in Figure \ref{fig:2}. The encoders of the normal TDL TDCs only have one stage of back-end circuits to find the 1-0 transition. In the arithmetic circuits, the adders process the selected local sums and the priority codes to obtain the final encoded transition position. Meanwhile, the en-flag signal indicates whether the hit occurs in the corresponding clock cycle.
There are two examples in Figure \ref{fig:3} showing how the encoder attains the transition positions. The final result has two parts: the base which is the product of piece width and sequence of the selected pieces, the offset which is the sum of local sums of the two selected pieces. The equation of encoding result is
\begin{equation}
\label{eq:b}
R=Base+Offset=P\cdot{W}+Sum
\end{equation}
In Equation \eqref{eq:b}, R represents the result of encoding, P represents the sequence of the first selected piece, W represents the width of the pieces (24 in Figure \ref{fig:3}), and Sum represents the sum of the local sums of the two adjacent selected pieces. All the calculation is completed by adders because the weights of the priority codes are fixed.

\subsubsection{Encoder for a half-length delay line TDC}

The waveform and raw data of the half-length delay line TDCs are shown in Figure \ref{fig:2}. To reduce the length of the delay line, half-length delay line TDCs utilize either rising or falling edge of the pulse signal. The difference between a half-length delay line TDC and a wave union TDC is that a half-length delay line TDC only deals with the first edge of the waveform. However, there are three kinds of situations that the encoder has to process. If traditional encoders are used for the half-length delay line TDC, the reduction of logic utilization will not be as large as the reduction of delay cells~\cite{kong2020resource}. 

The above-mentioned encoder used in a normal TDC can be easily adapted to half-length delay line TDCs. In this case, the flags and the MSB of the local sums of the selected pieces are different. The first flag is “1” and the second one is “0” when the 1-0 transition is registered. The flags are inverse for the 0-1 transition. As a result, the type of the captured edge can be distinguished in the final arithmetic circuits. Thus, the equation for the 1-0 transition is the same as for normal TDL TDCs, and the equation for the 0-1transition is
\begin{equation}
\label{eq:c}
R=P\cdot{W}+({W\cdot{2}-Sum}).
\end{equation}
The base part in the encoding result changes because the encoder counts "0"s in selected pieces in 0-1 transitions.

\subsubsection{Encoder for wave union TDC}

A wave union TDC has to deal with more than one edge in one sampled raw bin data~\cite{wu200810}. To solve this problem, more stages in the back-end circuit can be used to deal with more edges, as shown in Figure \ref{fig:5}. If the input data have more than one 1-0 or 0-1 transition, then there will be more than one "1" bits in the Sel array. In the first stage, the priority encoder finds the first edge and the first "1" bit in the Sel array is masked. The second stage can find the second one in the masked output of the first stage. It is the same in the subsequent back-end encoding stages. Because the type of the transitions delivered to the arithmetic circuits is clear, the arithmetic circuits handle the data according to their sequence. For example, the waveform in Figure \ref{fig:2} has two edges, and the first one is a rising edge and the second is a falling one. The final output is valid only when both edges are sampled and the encoder can indicate the number of transitions in the raw data.

\section{Implementation and test results}

To verify the newly presented encoding method, three kinds of TDL TDCs were implemented in a Xilinx low-end FPGA, which was a XC7A100TFGG484 device. Firstly, the normal TDL TDCs and corresponding encoders were realized. Then, the encoders were adapted for half-length delay line TDCs and wave union TDCs. Both double-edge and four-edge wave union TDCs were realized.

\begin{figure}[htb]
\centering 
\includegraphics[scale=0.3]{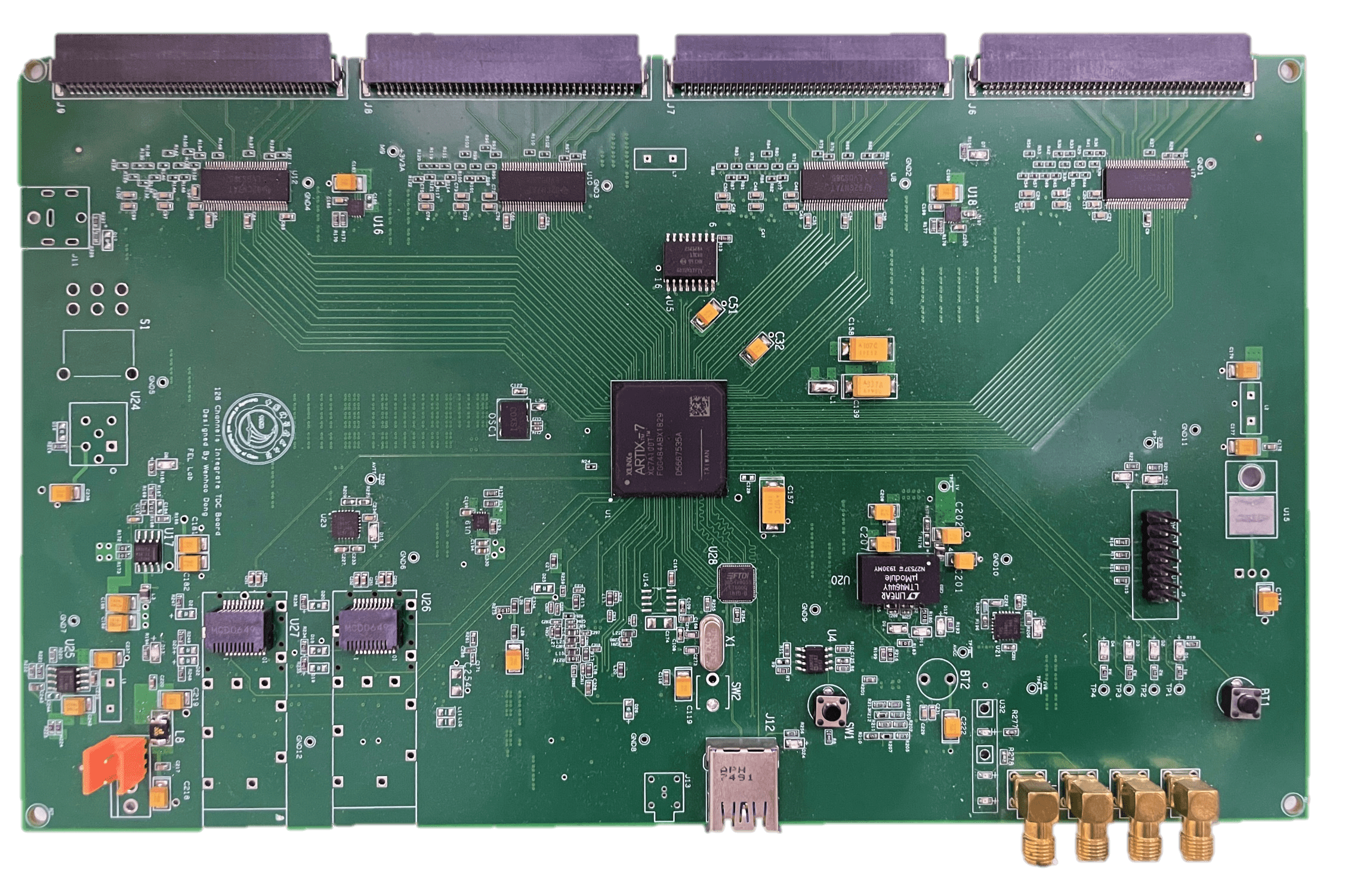}
\caption{\label{fig:6} The TDC prototype board.}
\end{figure}

A prototype board was designed with a high precision clock generator. The clock generator can provide different frequency clock signals while being under a picosecond of jitter. Therefore, the test results were hardly affected by the system clock jitter. An arbitrary waveform generator generated the pulse, and the pulse was transmitted to a matching tee joint. Two pulse waveforms with fixed delay were obtained using a pair of unequal length cables. As shown in Figure \ref{fig:6}, the signal was sent to the TDC board through an adapter board that was connected to the TDC board via an ERNI 154744 connecter. The digitalized timestamps were transferred to the PC via a USB port.

\subsection{Test results of the normal TDL TDC}

As shown in Figure \ref{fig:2}, the normal TDL TDC needs to capture the rising edge of the step signal in one clock period. This means that the whole delay of the delay line should be longer than the clock period. As for the FPGA-based TDCs, the delay line is realized by a carry chain that consists of carry logic in the FPGA. The carry logic in Xilinx Artix FPGA was CARRY4, which has four output ports. One CARRY4 corresponds to four delay bins. In the normal TDL TDCs, we used 54 CARRY4 elements to construct a delay line with 216 delay taps. The clock of the TDC block ran at 400 MHz. As a result, the input data of the encoder was 216-bit wide. The output of the 216-bit wide thermometer code input was an 8-bit wide binary code. One TDC channel consumed 254 LUTs and 449 registers. Bin-by-bin calibration was employed to get better time resolution~\cite{wu2010several}.

\begin{figure}[htb]
\centering 
\includegraphics[scale=0.45]{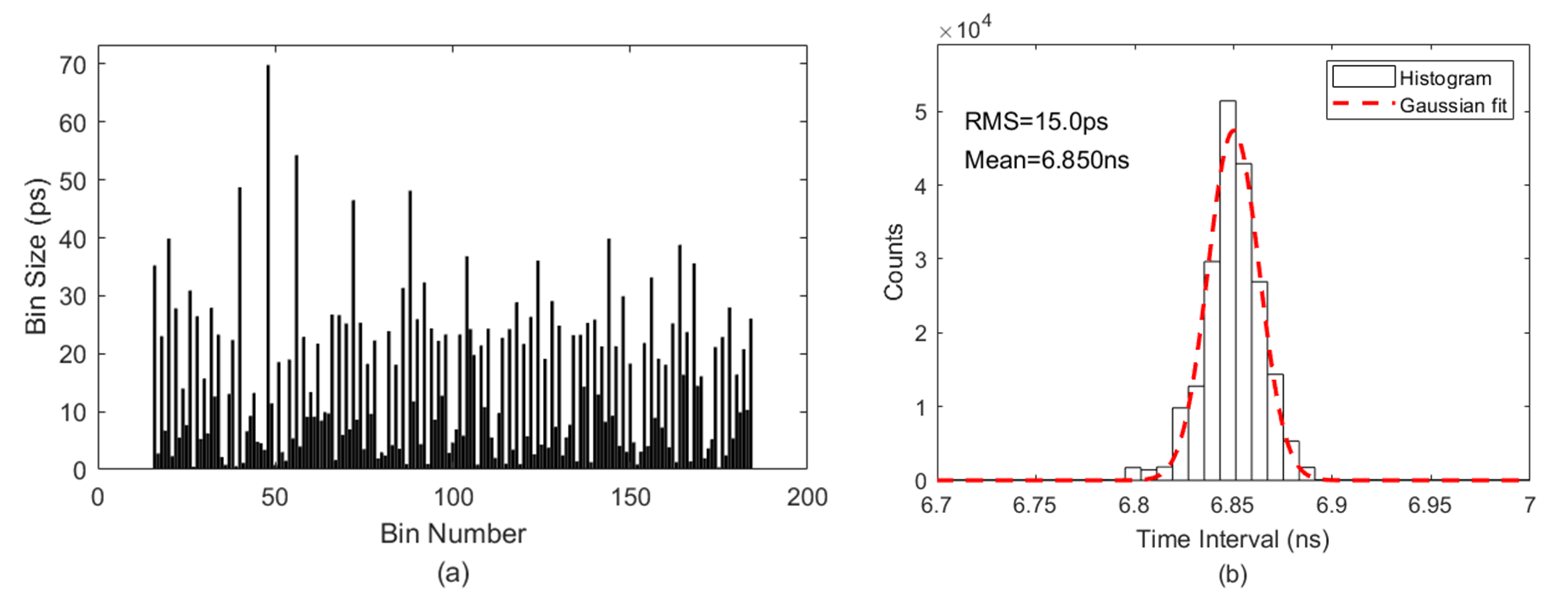}
\caption{\label{fig:7} Test results of the normal TDL TDC: (a) bin size of a single TDC channel, and (b) time resolution of two channel time-interval test after bin-by-bin calibration.}
\end{figure}

A statistical code density test was used to find the bin size of the TDL and the look-up table for bin-by-bin calibration. Figure \ref{fig:7} (a) shows the bin size test result of a single channel. In the test, 200,000 random hits were applied to the channel. The test showed that the average bin size was 14.8 ps. After calibration, the root mean square (RMS) value of the time-intervals evaluated by the “cable-delay test” of the two channels was 15.0 ps. This result is comparable to prior works, which means that the encoding method is effective in normal TDL TDCs.

\subsection{Test results of the half-length delay line TDC}

To reduce the length of the carry-chain, the half-length delay line TDCs take advantage of two edges of a pulse signal. In theory, this method can shorten the length of the delay-line by half. However, it is difficult to achieve this target in practice. Due to the uneven pulse width, more redundant delay cells had to be used to ensure that the pulse was sampled. In this module, the carry chain had 30 CARRY4 logic elements corresponding to 120 delay taps. One TDC channel consumed 156 LUTs and 248 registers. The sample region of the half length TDC is roughly equal to 192-bins in the normal TDCs. Because the width of the square pulse is slightly shorter than the length of the delay line, the sample region is slightly shorter than twice the delay line. The frequency of the system clock was 400 MHz, which was the same as the normal TDL TDC.

\begin{figure}[htb]
\centering 
\includegraphics[scale=0.45]{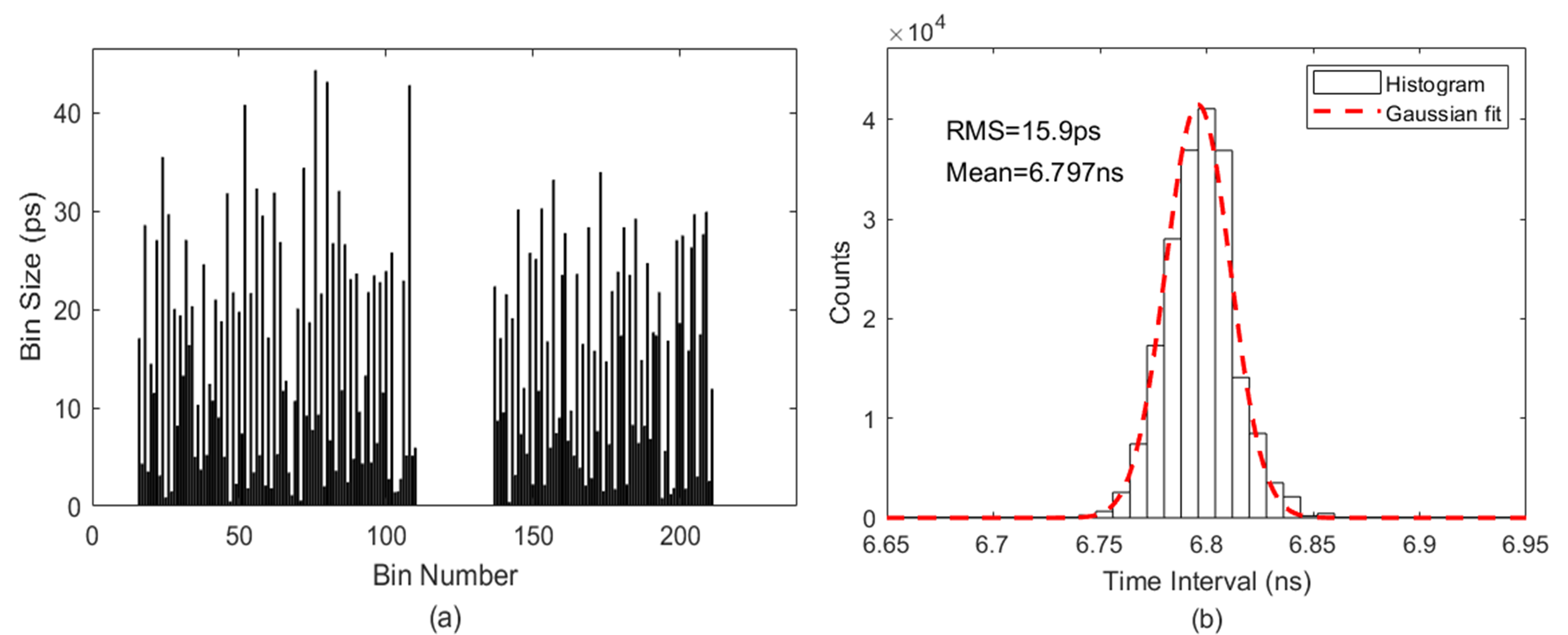}
\caption{\label{fig:8} Test results of the half-length delay line TDC: (a) bin size of a single TDC channel, and (b) time resolution of two channel time-interval test after bin by bin calibration.}
\end{figure}

Because the encoded fine code of the half-length delay line TDC had an extra bit to indicate the type of valid transition, the 120-bin carry chain also had an 8-bit binary fine code to be calibrated. As shown in  Figure \ref{fig:8}, the encoded numbers were divided into two parts by the flag bit. The average bin size of the half-length delay line TDC in this channel was 14.6 ps, and the time resolution between the two channels was 15.9 ps.

\subsection{Test results of wave union TDCs}

To explore the logic utilization of wave union TDCs using the resource-saving method, double-edge and four-edge wave union TDCs were implemented in FPGA. The waveform and the raw data are shown in Figure \ref{fig:2}. In this case, a double-edge wave union TDC was realized on a 288-bin carry chain, and a four-edge wave union TDC was realized on a 360-bin carry chain. The clock frequencies were the same at 400 MHz. The circuits to obtain the local sums and flags were the same as in the normal and half-length delay line TDCs. The back-end of the encoder had more stages of priority encoders, buffers, and multiplexers to deal with more edges. The arithmetic circuits in each stage calculated the edge position by the result from the priority encoder and its corresponding local sums. Finally, the positions of the two or four edges were added together. The double-edge wave union TDCs consumed 438 LUTs and 761 registers per channel, and the four-edge wave union TDCs consumed 706 LUTs and 1,263 registers per channel. The pulse generated by the launcher must be longer than 24 bins’ delay because the pieces in the pre-encoded cells were 24-bits wide. Finally, 9-bit wide binary code results in double-edge wave union TDCs and 11-bit wide results in four-edge TDCs were obtained. 

\begin{figure}[htb]
\centering 
\includegraphics[scale=0.45]{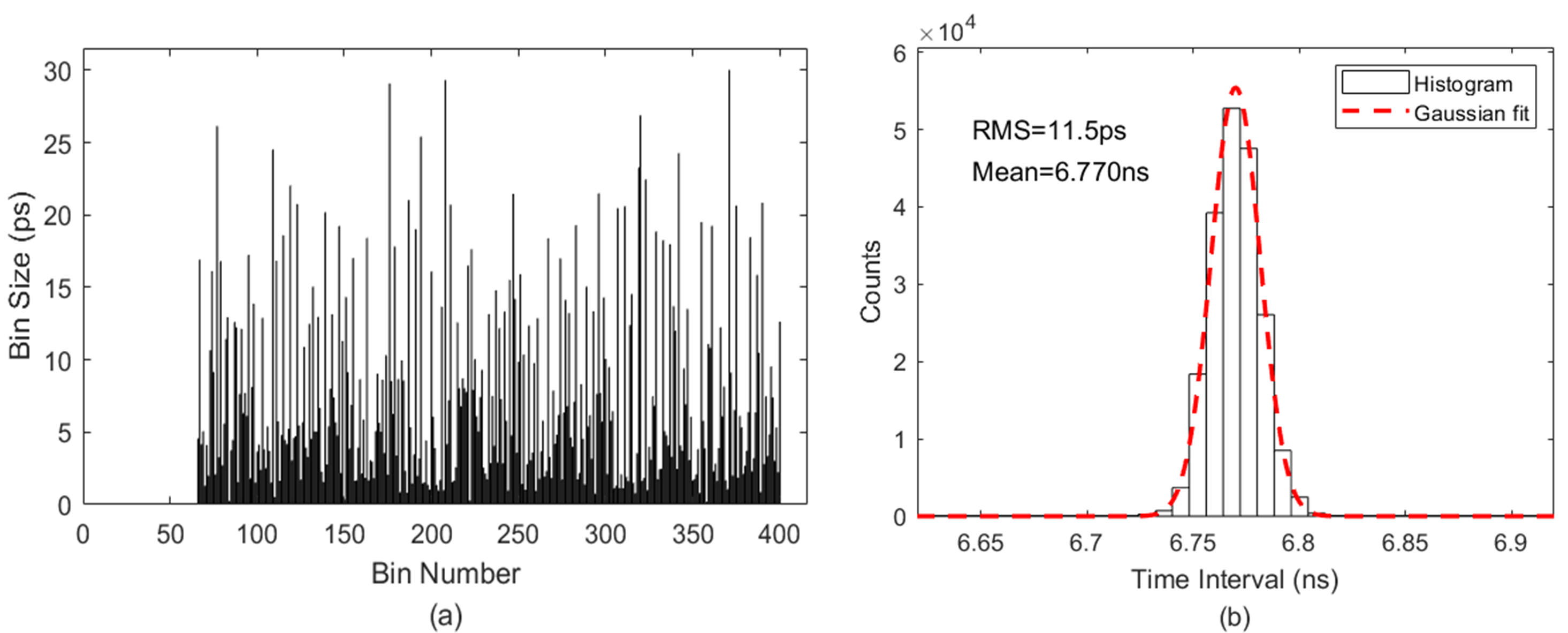}
\caption{\label{fig:9} Test results of double-edge wave union TDC: (a) bin size of a single TDC channel, and(b) time resolution of the two channel time-interval test after bin-by-bin calibration.}
\end{figure}

The smallest number of the fine code was not zero because the fine code was the sum of two transitions’ locations. The bin size test result shown in Figure \ref{fig:9}(a) indicates that the fine code ranged from 66 to 403. Thus, the average bin size of double-edge wave union TDC was 7.4 ps, and the time resolution of the two tested channels was 11.5 ps.

\begin{figure}[htb]
\centering 
\includegraphics[scale=0.45]{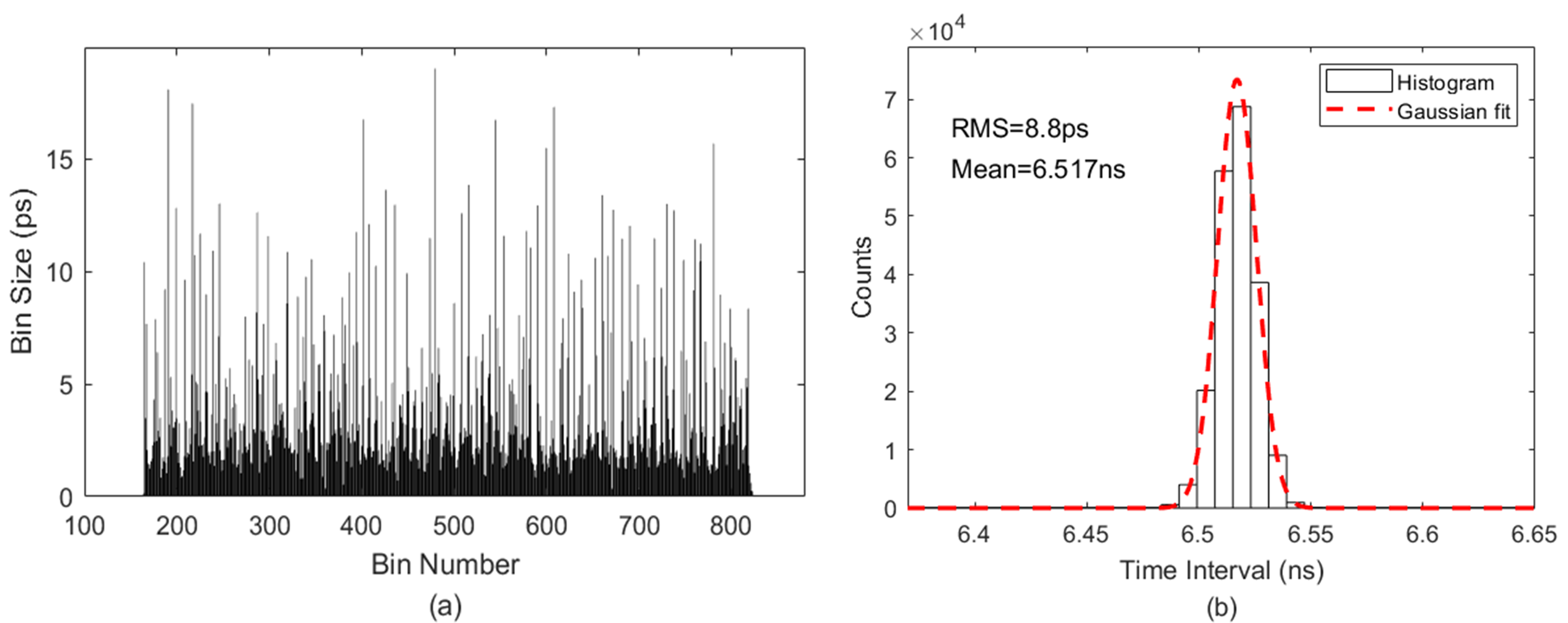}
\caption{\label{fig:10} Test results of four-edge wave union TDC: (a) bin size of a single TDC channel, and (b) time resolution of two channel time-interval test after bin-by-bin calibration.}
\end{figure}

The smallest number of fine codes of the four-edge wave union TDCs was also not zero. The bin size test result shown in Figure \ref{fig:10}(a) indicates that the fine code ranged from 164 to 825. Thus, the average bin size of four-edge wave union TDC was 3.8 ps, and the time resolution of the two tested channels was 8.8 ps.

\section{Discussion}

\subsection{Resource utilization}

Table \ref{tab:i} provides comparisons of the resource utilization of three kinds of TDCs in some recent works. In the table, ‘$\sim$’means that the parameter is not mentioned in the text of the paper, but deduced from figures with an approximate value. Most of the TDCs listed in Table \ref{tab:i} are implemented on Xilinx 7 series FPGAs, which are the same as the ones in this paper. As for the work in ref ~\cite{xie2020wave}, a Xilinx UltraScale Kintex FPGA was used, which has the same architecture in LUTs and registers as 7 series.

However, it is difficult to compare these encoders directly. The clock frequency, sensitivity to bubbles and dead time all have impacts on resource utilization. And all these TDCs have different length of delay line. Considering that all the works are implemented in similar FPGA devices, the usages of LUTs and registers divided by the number of total taps in each TDL are listed in the table, which can be seen as two simplified (though not precise) indexes for a rough comparison.

As for the normal TDL TDCs, Xie’s encoder, which is the best one among the previous works, used 83\% more LUTs than this paper ~\cite{xie2020multi}. It ran at a clock frequency of about 250M Hz, which is much lower than most of other works (about 400 MHz or higher). Another TDCs in Zheng’s paper only used half of the four outputs of the CARRY4 element, and the clock frequency is also 250 MHz ~\cite{zheng2017low}, was lower than other works. It should be pointed out that for a lower clock frequency, the pipeline stages of the encoder could be reduced, which results in the usage of fewer registers.

\begin{table}[htb]
\centering
\caption{\label{tab:i}  Resource utilization of a single channel.}
\smallskip
\begin{tabular}{|cccccc|c|}
\hline
\multicolumn{1}{|c|}{}                     & \multicolumn{1}{c|}{Taps}      & \multicolumn{1}{c|}{LUTs} & \multicolumn{1}{c|}{Registers} & \multicolumn{1}{c|}{LUTs/ Taps} & \multicolumn{1}{c|}{Registers/Taps} & \begin{tabular}[c]{@{}c@{}}Device\\ Family\end{tabular}            \\ \hline
\multicolumn{7}{|c|}{Normal TDL TDCs}                                                                                                                                                                                                           \\ \hline
\multicolumn{1}{|c|}{TNS’17~\cite{zheng2017low}}               & \multicolumn{1}{c|}{140}       & \multicolumn{1}{c|}{668}        & \multicolumn{1}{c|}{387}            & \multicolumn{1}{c|}{4.77}       & \multicolumn{1}{c|}{2.76}           & Artix-7           \\ \hline
\multicolumn{1}{|c|}{TNS’17~\cite{wang20173}}               & \multicolumn{1}{c|}{200}       & \multicolumn{1}{c|}{505}        & \multicolumn{1}{c|}{1109}           & \multicolumn{1}{c|}{2.525}      & \multicolumn{1}{c|}{5.545}          & Kintex-7          \\ \hline
\multicolumn{1}{|c|}{TNS’17~\cite{qin20171}}               & \multicolumn{1}{c|}{$\sim$170} & \multicolumn{1}{c|}{848}        & \multicolumn{1}{c|}{/}              & \multicolumn{1}{c|}{4.99}       & \multicolumn{1}{c|}{/}              & Virtex-7          \\ \hline
\multicolumn{1}{|c|}{TIE’19~\cite{chen2018multichannel}}               & \multicolumn{1}{c|}{$\sim$380} & \multicolumn{1}{c|}{1145}       & \multicolumn{1}{c|}{1916}           & \multicolumn{1}{c|}{3.01}       & \multicolumn{1}{c|}{5.04}           & Virtex-7          \\ \hline
\multicolumn{1}{|c|}{EBCCSP’20~\cite{xie2020multi}}            & \multicolumn{1}{c|}{$\sim$380} & \multicolumn{1}{c|}{822}        & \multicolumn{1}{c|}{1219}           & \multicolumn{1}{c|}{2.16}       & \multicolumn{1}{c|}{3.21}           & Virtex-7          \\ \hline
\multicolumn{1}{|c|}{\textit{This   Work}} & \multicolumn{1}{c|}{216}       & \multicolumn{1}{c|}{451}        & \multicolumn{1}{c|}{805}            & \multicolumn{1}{c|}{1.18}       & \multicolumn{1}{c|}{2.08}           & Artix-7           \\ \hline
\multicolumn{7}{|c|}{The half-length delay line TDCs}                                                                                                                                                                                           \\ \hline
\multicolumn{1}{|c|}{I2MIC’20~\cite{kong2020resource}}             & \multicolumn{1}{c|}{96}        & \multicolumn{1}{c|}{418}        & \multicolumn{1}{c|}{852}            & \multicolumn{1}{c|}{4.35}       & \multicolumn{1}{c|}{8.88}           & Virtex-7          \\ \hline
\multicolumn{1}{|c|}{\textit{This   Work}} & \multicolumn{1}{c|}{120}       & \multicolumn{1}{c|}{229}        & \multicolumn{1}{c|}{388}            & \multicolumn{1}{c|}{1.3}        & \multicolumn{1}{c|}{3.23}           & Artix-7           \\ \hline
\multicolumn{7}{|c|}{Double-edge Wave Union TDCs}                                                                                                                                                                                               \\ \hline
\multicolumn{1}{|c|}{arXiv’20~\cite{xie2020wave}}             & \multicolumn{1}{c|}{480}       & \multicolumn{1}{c|}{1349}       & \multicolumn{1}{c|}{1480}           & \multicolumn{1}{c|}{2.81}       & \multicolumn{1}{c|}{3.83}           & \begin{tabular}[c]{@{}c@{}}UltraScale \\ Kintex\end{tabular} \\ \hline
\multicolumn{1}{|c|}{TNS’19~\cite{wang20193}}               & \multicolumn{1}{c|}{216}       & \multicolumn{1}{c|}{1085}       & \multicolumn{1}{c|}{2164}           & \multicolumn{1}{c|}{5.02}       & \multicolumn{1}{c|}{10.02}          & Kintex-7          \\ \hline
\multicolumn{1}{|c|}{\textit{This   Work}} & \multicolumn{1}{c|}{288}       & \multicolumn{1}{c|}{438}        & \multicolumn{1}{c|}{761}            & \multicolumn{1}{c|}{1.52}       & \multicolumn{1}{c|}{2.91}           & Artix-7           \\ \hline
\multicolumn{7}{|c|}{Four-edge Wave Union TDCs}                                                                                                                                                                                                 \\ \hline
\multicolumn{1}{|c|}{TNS’19~\cite{wang20193}}               & \multicolumn{1}{c|}{288}       & \multicolumn{1}{c|}{1410}       & \multicolumn{1}{c|}{2732}           & \multicolumn{1}{c|}{4.9}        & \multicolumn{1}{c|}{9.49}           & Kintex-7          \\ \hline
\multicolumn{1}{|c|}{\textit{This Work}}   & \multicolumn{1}{c|}{360}       & \multicolumn{1}{c|}{706}        & \multicolumn{1}{c|}{1263}           & \multicolumn{1}{c|}{1.96}       & \multicolumn{1}{c|}{3.51}           & Artix-7           \\ \hline
\end{tabular}
\end{table}
The greatest advantage of the encoding method presented in this paper is that the logic utilization of NTH2B encoding can be largely reduced. The encoder of half-length delay line TDC saves half of the resources compared with the original work~\cite{kong2020resource}. Compared with the presented encoder for wave union TDC, the advantage of our encoding method is greater than TH2B encoding~\cite{chen2018multichannel,xie2020wave}. This encoder uses 83\% more LUTs than encoder in this paper, and the consumption of register is 32\%  more. Furthermore, the encoding method proposed in prior work~\cite{knittel2019novel} only deals with the double-edge wave union TDCs. The best results in Table \ref{tab:i} used this method and only realized double-edge wave union TDCs~\cite{xie2020wave}.  When the TDL has more edges, the proposed encoding method will have greater advantages~\cite{wang20193}. Because it can locate more transitions by increasing the stages of back-end circuits without changing the original function blocks, it is convenient to adapt our encoding method for TDL, which has more edges with less increased resource usage.

\subsection{Dead time}

The resource-saving encoding method described above is implemented using pipeline techniques so that the dead time of encoders is one clock period. However, the flags of the encoder in two adjacent periods are used to generate the write-enable signal of the FIFO in the TDC sub-system. Therefore, the dead times of the whole TDC circuits are two clock periods. The dead time is acceptable in most cases as the clock frequency of the TDC system clock is very high.

\subsection{Sensitivity to bubbles}

The balance of logic utilization, dead time and sensitivity to bubbles should be made for different applications. The encoding block in Figure \ref{fig:4} divided the raw data into 24-bit wide pieces. Assuming that 0 and 1 in the blurring area of bubbles are roughly uniform, as observed in previous work~\cite{wang20164}, the encoder can handle 16-bit wide bubbles in 1-to-0 transition in a worst-case scenario. The worst case is that the transition locates in the second selected piece and the first piece has no bubbles. In that case, if the length of bubbles is to large, two adjacent piece will not cover all blurring area. Because the threshold is 16 which means the MSB is easier turn to zero in this example, this encoding circuit is more sensitive to bubbles of 1-to-0 transitions.

The bubble depth is 4 bit in 40nm Xilinx FPGA~\cite{jiao2021resource}. Then it increases to 5 bit in 28nm device and more than 10 bits in 20nm and 14nm devices~\cite{song2018256,wang20173,wu2021design}. The 24bit wide piece in this paper is capable for the bubble problems in the Artix-7 FPGA which is on the 28nm process. If the encoding method is employed on TDCs with higher bubble depth, wider pieces can solve the problem.

\section{Conclusion}

In this paper, a novel resource-saving, low dead time encoding method for TDL TDCs was proposed. This encoding method was inspired by the divide-and-conquer strategy, and it combines two encoding methods, sum encoding and priority encoding. This method not only greatly reduces the use of logic resources but also solves the high depth bubble problem. And three kinds of TDCs have been realized in the Artix 7 Device. Both acceptable resolutions and improvements on resource efficiency are attained. With this method, the TDL TDCs will have greater prospects in emerging applications. 

\acknowledgments

This work was supported by the Strategic Priority Research Program of Chinese Academy of Sciences, Grant No. XDC07020300.
\bibliographystyle{abbrv}
\bibliography{ref1}

\end{document}